# INVESTIGATION OF PHONON DYNAMICS OF PEROVSKITE MULTIFERROIC MANGANITES: $R$MnO$_3$ (R=Tb, Dy, Ho)

**Renu Choithrani[1], Mala N Rao[2], S. L. Chaplot[2], N. K. Gaur[1] and R. K. Singh[3]**
[1]Department of Physics, Barkatullah University, Bhopal - 462026, India
[2]Solid State Physics Division, Bhabha Atomic Research Centre, Mumbai - 400085, India
[3]School of Basic Sciences, MATS University, Raipur – 492002, India

## Abstract

We have used a shell model to study the phonon dynamics of multiferroic manganites $R$MnO$_3$ (R= Tb, Dy, Ho). The calculated phonon dynamical properties, crystal structure, Raman frequencies and specific heat are found to be in good agreement with the available experimental data. Besides, the phonon density of states, elastic constants and phonon dispersion curves along high symmetry directions ($\Sigma$, $\Delta$ and $\Lambda$) have also been computed. A zone-center imaginary A$_u$ mode is revealed in these phonon dispersion curves, which indicates the occurrence of metastability of the perovskite phase. The Gibbs free energy values of orthorhombic phase, when compared with those of hexagonal phase, indicate the possibility of coexistence of these two phases of these multiferroic manganites under ambient conditions.

# 1. INTRODUCTION

Interest in multiferroic manganites has increased rapidly, both experimentally and theoretically, due to the strong interrelation between their structures, electronic and magnetic properties. It is noted from the literature that most of the investigations done on them are mainly on $LaMnO_3$, $PrMnO_3$ and $NdMnO_3$. Despite intense experimental investigations, there is not much published information on the phonon dynamical properties of $RMnO_3$ (Tb, Dy, Ho). The compounds with perovskite-type structure can be treated as model materials because of the richness in their physical properties and relatively simple structure. These materials have several technological applications such as computer memories, pyroelectric sensors, piezoelectric transducers, and multilayered capacitors. $TbMnO_3$ (TMO) has recently attracted particular attention in the field of multiferroic materials [1]. For this reason, our aim in the present work is to investigate in detail the phonon dynamical properties of these systems.

Recently, gigantic magnetoelectric and magnetocapacitance effects have been found in the orthorhombic perovskite $TbMnO_3$. In $TbMnO_3$, a strong coupling between the magnetic-order and polarization-order parameters suggests that the ferroelectricity might be induced by incommensurate magnetism [2, 3].

Martin Carron *et al*. [4] have studied orthorhombic (o-)$RMnO_3$ (R = Pr, Nd, Eu, Tb, Dy, Ho) manganites for their Raman phonons as a function of the rare earth ions and temperature. They have correlated the frequencies of the three most intense modes of these orthorhombic compounds with some structural parameters like Mn-O bond distances, octahedral tilt angle and Jahn-Teller distortion. Later on, the polarized Raman spectra for a series o-$RMnO_3$ (R = Pr, Nd, Eu, Gd, Tb, Dy, Ho) have been observed by Iliev *et al*. [5] where they have assigned the observed frequencies, as well as studied the basic distortions responsible for the mode activation. Their study shows that the variation of lattice distortion with radius of rare earth atoms affects significantly both the phonon frequencies and the shape of some of Raman modes.

The crystal structure of o-$RMnO_3$ (R = Tb, Dy, Ho) exhibits distorted perovskite structure with space group *Pnma* at room temperature [5-7]. Their magnetic structures have also been studied by means of x-ray diffraction (XRD) measurements [8], as well as neutron diffraction [9]. The magnetization of $TbMnO_3$, shows two sharp peaks around 27 and 42 K. In case of $HoMnO_3$, there are magnetic transitions below 60 K as observed by Tachibana et al [10], and by Munoz et al. [11]; a sharp peak around $T_N \sim 40$ K marks antiferromagnetic (AF) transition into the incommensurate (IC) phase, additional anomaly is observed at $T'_N \sim 5\text{-}15$ K below $T_N$ and this feature is ascribed to a lock-in-transition to the E-type structure.

HoMnO$_3$ has been known to exist in the hexagonal phase (Space group P6$_3$cm) [12], but can be transformed into its orthorhombic phase by annealing under high pressure. For DyMnO$_3$, the hexagonal phase may be obtained by quenching from high temperature (≥ 1600º C), as also through epitaxial growth [13, 14]. Further, the hexagonal TbMnO$_3$ metastable phase epitaxially stabilized in thin-film form, displays nearly twenty times larger remnant polarization with the ferroelectric ordering temperature shifted to 60 K (compared to 27 K in the bulk orthorhombic phase [15].

Recently, a modified rigid ion model (MRIM) has been used by some of us [16] to elucidate the cohesive, thermal and thermodynamic properties of pure and doped perovskite manganites. Earlier, a rigid ion model (RIM) has been used for successful description of the lattice dynamics and cohesion of monovalent metal nitrates [17], and divalent metal carbonates [18]. Also, the lattice dynamical investigations of (Sm, Eu, Gd) MnO$_3$ [19], LaMnO$_3$ [20] and YMnO$_3$ [21] have been reported by using a seven parameter shell model [22-26] based on a transferable interatomic potential.

In the present study, we have used this lattice dynamical shell model [19] for describing the phonon dynamical and related properties of the multiferroic manganites $R$MnO$_3$ (R=Tb, Dy, Ho) and computed their crystal structure, phonon density of states, phonon dispersion relations, Raman frequencies, elastic constants, Debye temperature, specific heat and Gibbs free energy. The symmetry vectors obtained through the group theoretical analysis for *Pnma* space group along the zone centre (Γ) point and along the high symmetry directions have been employed to classify the phonon frequencies obtained into their irreducible representations. We have used the present shell model to compute the phonon dispersion curves for RMnO$_3$ (R= Tb, Dy, Ho) along three major symmetry (∑, Δ and Λ) directions. The essential formalism of the present shell model and method of computations have been presented in the following section.

## 2. SHELL MODEL COMPUTATIONS

The interatomic potential of the shell model consists of the long-range Coulomb and the short-range interactions [19] and employs the effective ionic charge and effective radius associated with the atoms, as the adjustable parameters. The effect of polarizability of the oxygen ions in this model has been included by considering a massless shell with charge Y linked to the core of the ions by harmonic interactions with a shell-core force constant K.

The shell model parameters are evaluated in such a way that the forces on each atom and the internal stresses in the crystal vanish and they reproduce the crystal structure close to that observed by the neutron diffraction experiments. These computations are performed using the DISPR [24] program. The optimized parameters for $R$MnO$_3$ (R=Tb, Dy, Ho) thus obtained are listed in Table 1. The phonon dispersion curves (PDCs) for the orthorhombic phase have been computed using the equilibrium structure at the minimum of the potential energy.

The phonon density of states is given by [22, 26]

$$g(\omega) = N\int_{BZ} \sum \delta[\omega - \omega_j(q)]dq, \quad (1)$$

with $N$ as a normalization constant such that $\int g(\omega)d\omega = 1$ and $g(\omega)d\omega$ is the ratio of the number of eigenstates in the frequency interval ($\omega$, $\omega + d\omega$) to the total number of eigenstates. Here, $\omega_j(q)$ is the phonon frequency of jth normal mode of the wave vector q. These density of states have been used to evaluate the specific heat at constant volume [22, 26]

$$C_v(T) = k_B \int \left(\frac{\hbar\omega}{k_BT}\right)^2 \frac{e^{(\hbar\omega/k_BT)}}{[e^{(\hbar\omega/k_BT)} - 1]^2} g(\omega)d\omega.$$

(2)

at different temperatures (*T*). Here, $k_B$ and $h$ are the Boltzmann and Planck constants, respectively. The other symbols are the same as defined elsewhere [21, 26]. The procedure adopted to compute these lattice dynamical and allied properties is prescribed in detail elsewhere [22-26].

Besides, the group-theoretical analysis for the perovskite manganites shows that the phonon modes at the zone center point (Γ) may be classified as [20-23] $7A_g+7B_{2g}+5B_{1g}+5B_{3g}+10B_{1u}+10B_{3u}+8A_u+8B_{2u}$. Out of these 60 phonon modes, 24($7A_g+5B_{1g}+7B_{2g}+5B_{3g}$) are Raman active, 25 ($9B_{1u}+7B_{2u}+9B_{3u}$) are infrared active, 8 ($8A_u$) are silent, and 3 ($B_{1u}+B_{2u}+B_{3u}$) are acoustic modes. The following group-theoretical classification has been obtained along the three high-symmetry directions:

$\Sigma : 17\Sigma_1+13\Sigma_2+13\Sigma_3+17\Sigma_4,$

$\Delta : 15\Delta_1+15\Delta_2+15\Delta_3+15\Delta_4, \quad (3)$

$\Lambda : 17\Lambda_1+13\Lambda_2+13\Lambda_3+17\Lambda_4.$

The dynamical matrix has been diagonalized along these symmetry directions and the normal modes have been classified into different representations to obtain the phonon-dispersion relations in $RMnO_3$ (R = Tb, Dy, Ho). The computed results are presented and discussed below.

## 3. RESULTS AND DISCUSSION

### 3.1 CRYSTAL STRUCTURE

We have applied the DISPR program [24] (developed at BARC) to compute the optimized potential parameters and the atomic coordinates as listed in Tables 1 and 2, respectively. These parameters yield nearly vanishing forces on any atom. The values of these parameters have been used to compute the eigenvalues of the dynamical matrix corresponding to the present shell model.

The crystal structure of o-RMnO$_3$ (R = Tb, Dy, Ho) has been calculated by minimizing the Gibbs free energy with respect to the structural parameters. The multiferroic perovskite manganites RMnO$_3$ (R = Tb, Dy, Ho) (with space group *Pnma*) have an orthorhombic structure with four formula units per unit cell [6-7]. It has five atoms in the asymmetric unit with R and O occupying crystallographic sites 4(c), except for one oxygen atom, which is at the general position 8 (d) and Mn is at 4 (a) position. This structure contains a network of corner sharing MnO$_6$ octahedra and belongs to the family of rotationally distorted perovskites with Glazer's notation (*a−b+a−*). It can be obtained from the simple perovskite structure by two consequent rotations of MnO$_6$ octahedra, namely, (i) around (010) direction of the cubic perovskite (*y* axis in *Pnma* ) and (ii) around (101) direction of cubic perovskite (*x* axis in *Pnma*). The structure has two oxygen ions O1 and O2 with different site symmetry. The O1 ion lies in the symmetry plane. In *Pnma* space group, the R and O1 atoms have the site symetry *Cs*. The site symmetry of Mn ion is *Ci* and that of O2 is $C_1$.

**Table 1:** The potential parameters obtained for RMnO$_3$ (R=Tb, Dy, Ho) using the shell model.

|  | Tb | Dy | Ho | Mn | O |
|---|---|---|---|---|---|
| Z(k) (e) | 2.75 | 2.75 | 2.75 | 1.00 | -1.25 |
| R(k) Å | 2.0171 | 2.016 | 2.00 | 1.07 | 1.71 |
|  |  |  |  |  | Y(O)=-1.8 e  K(O)=183 eV Å$^2$ |

The values of the calculated lattice parameters (a, b, c), atomic coordinates (x, y, z), unit cell volumes (V) and strain parameter (s(=2(a-c)/(a+c)) are listed in Table 2 for orthorhombic manganites RMnO$_3$ (R=Tb, Dy, Ho) and compared with the available experimental neutron diffraction data [6, 7]. It is found that our calculated results are in fairly good agreement with their corresponding experimental data. The calculated lattice parameters (a, b, c) differ only by 0.8% on an average from the experimentally observed data [6,7] for RMnO$_3$ (R=Tb, Dy, Ho). The calculated values of the unit cell volumes (V) are also in good agreement with experimental data [6, 7]. The decreasing trend of V with decreasing order of ionic radius ($r_R$) exhibited by both experimental and our theoretical results are similar as is seen from Table 2.

**Table 2:** Comparison of the experimental data [6, 7] (at 295 K) with calculated structural parameters (at 0 K) for $RMnO_3$ (R=Tb, Dy, Ho). For the space group *Pnma*, (Tb, Dy or Ho), Mn, O1 and O2 atoms are located at (x, 0.25, z), (0, 0, 0), (x, 0.25, z) and (x, y, z), respectively and their symmetry equivalent positions.

|  | $TbMnO_3$ | | $DyMnO_3$ | | $HoMnO_3$ | |
|---|---|---|---|---|---|---|
|  | Expt.[6] | Calc. | Expt.[7] | Calc. | Expt.[6] | Calc. |
| a (Å) | 5.8384(1) | 5.817 | 5.84480(25) | 5.816 | 5.8354(1) | 5.779 |
| b (Å) | 7.4025(1) | 7.438 | 7.37885(34) | 7.436 | 7.3606(1) | 7.441 |
| c (Å) | 5.29314(5) | 5.228 | 5.28016(22) | 5.227 | 5.2572(1) | 5.213 |
| V(Å$^3$) | 228.760(6) | 226.2 | 227.722(17) | 226.1 | 225.807(4) | 224.2 |
| S | 0.0980 | 0.107 | 0.1015 | 0.107 | 0.1043 | 0.103 |
| R(Tb/Dy/Ho) | | | | | | |
| X | 0.0824(3) | 0.094 | 0.08139(27) | 0.094 | 0.0839(3) | 0.093 |
| Z | 0.9831(4) | 0.985 | 0.9820(4) | 0.985 | 0.9825(3) | 0.984 |
| O1 | | | | | | |
| X | 0.4667(4) | 0.442 | 0.4648(8) | 0.442 | 0.4622(4) | 0.441 |
| Z | 0.1038(5) | 0.134 | 0.1017(9) | 0.134 | 0.1113(4) | 0.136 |
| O2 | | | | | | |
| X | 0.3262(3) | 0.331 | 0.3249(6) | 0.331 | 0.3281(3) | 0.330 |
| Y | 0.0510(2) | 0.067 | 0.0518(5) | 0.067 | 0.0534(2) | 0.068 |
| Z | 0.7039(4) | 0.686 | 0.7013(6) | 0.686 | 0.7013(3) | 0.686 |

### 3.2 ELASTIC CONSTANTS

We have calculated the elastic constants ($C_{11}$, $C_{22}$, $C_{33}$, $C_{44}$, $C_{55}$ and $C_{66}$) for $RMnO_3$ (R = Tb, Dy, Ho) through a computation of the acoustic wave velocities along different symmetry directions ($\Sigma$, $\Delta$ and $\Lambda$) and listed them in Table 3. These calculated values of the elastic constants could not be compared due to the lack of experimental data on them. Although these values are of only academic interest at present but they will serve as a guide to the experimental workers in future.

**Table 3**: The elastic constants (GPa) of RMnO$_3$ (R=Tb, Dy, Ho).

|        | $C_{11}$ | $C_{22}$ | $C_{33}$ | $C_{44}$ | $C_{55}$ | $C_{66}$ |
|--------|------|------|------|------|------|------|
| TbMnO$_3$ | 170 | 150 | 197 | 48 | 73 | 77 |
| DyMnO$_3$ | 170 | 151 | 196 | 48 | 73 | 76 |
| HoMnO$_3$ | 178 | 143 | 201 | 43 | 74 | 76 |

### 3.3 RAMAN AND INFRARED SPECTRA

We report the calculated Raman frequencies for o-RMnO$_3$ (R = Tb, Dy, Ho) in Table 4 along with the available experimental data [5, 27]. These phonon frequencies calculated by us are in good agreement with the observed Raman data [5, 27]. A comparison of the experimental data [5, 27] with other theoretical results obtained from the shell-model computations performed by Iliev et. al. [5] shows that the average deviation of the calculated Raman frequencies is 11.8%, whereas the present shell model has only 7% deviation, on an average. Also, Iliev et.al. [5] have used a much more complex shell model with twenty parameters, while our shell model has only seven parameters. The present shell model has revealed all 24 Raman active modes (7A$_g$ + 5B$_{1g}$ + 7B$_{2g}$ + 5B$_{3g}$) of vibrations as listed in Table 4. It is also noticed from this Table that the calculated Raman frequencies, particularly, for (7A$_g$+5B$_g$) modes are generally in good agreement with both the available experimental data [5, 27] in almost all three orthomanganites RMnO$_3$ (R=Tb, Dy, Ho). This feature emphasizes the conjective that the ionic interactions play a significant role in describing the vibrational properties of the present system of manganites.

The present shell model has also been used to calculate 25 infrared frequencies along with LO-TO splitting as depicted in Fig. 1. However, these results could not be compared due to the lack of experimental data.

**Table 4:** Comparison of Calculated and Experimental [5, 27] Raman-active Phonon Frequencies for RMnO$_3$ (R=Tb, Dy, Ho).

| Raman data | TbMnO$_3$ | | | DyMnO$_3$ | | HoMnO$_3$ | |
|---|---|---|---|---|---|---|---|
| | Calc. | Expt. [5] | Expt. [27] | Calc. | Expt. [5] | Calc. | Expt. [5] |
| **A$_g$** | 108 | - | - | 106 | - | 106 | - |
| | 146 | - | - | 145 | - | 144 | - |
| | 236 | 281 | 284 | 236 | - | 238 | 288 |
| | 273 | 315 | 316 | 273 | 320 | 275 | 324 |
| | 338 | 378 | 384 | 338 | 386 | 339 | 396 |
| | 446 | 489 | 494 | 446 | 492 | 452 | 499 |
| | 510 | 509 | 511 | 510 | 513 | 513 | 520 |
| **B$_{2g}$** | 101 | - | - | 100 | - | 99 | - |
| | 199 | - | - | 196 | - | 196 | - |
| | 287 | - | 309 | 288 | - | 290 | - |
| | 350 | 331 | 336 | 349 | 336 | 354 | - |
| | 498 | 474 | 478 | 498 | 478 | 503 | 481 |
| | 524 | 528 | 531 | 524 | 534 | 527 | 537 |
| | 573 | 612 | 614 | 573 | 614 | 577 | 615 |
| **B$_{1g}$** | 132 | - | | 131 | - | 125 | - |
| | 206 | - | | 206 | - | 206 | - |
| | 268 | - | | 267 | - | 267 | - |
| | 296 | - | | 297 | - | 298 | - |
| | 530 | - | | 530 | - | 531 | - |
| **B$_{3g}$** | 102 | - | | 101 | - | 95 | - |
| | 249 | - | | 250 | - | 251 | - |
| | 284 | - | | 284 | - | 285 | - |
| | 454 | - | | 454 | - | 455 | - |
| | 504 | - | | 505 | - | 505 | - |

## 3.4 PHONON DENSITY OF STATES

The total phonon density of states (PDOS) have been computed by using Eq. 1 and the results are depicted in Fig. 2 for RMnO$_3$(R=Tb, Dy, Ho), respectively. These total PDOS span the spectral range from 0 to 80 meV. Also, they have shown broad structures in 10 to 50 meV range. It is also interesting to note from Fig. 2 that the peaks around 65 and 72 meV in total PDOS can be attributed to vibrations of oxygen ions.

The partial phonon density of states of individual atoms (Tb, Dy, Ho), Mn, O1 and O2 have also been computed and shown in Fig. 2. It is noticed from these figures that (Tb, Dy or Ho) atoms contribute, generally, up to 40 meV, and the Mn atoms contribute mainly up to 30 meV.

## 3.5 SPECIFIC HEAT

The calculated density of states has been used to evaluate the specific heat at constant volume (C$_V$) as a function of temperature (T) using Eq. 2. These values have been computed for RMnO$_3$ (R = Tb, Dy, Ho) up to the temperature range of 1000 K and displayed in Fig. 3 and compared with the available experimental results [8, 10]. It is noticed from Fig. 3 (a) that the calculated specific heat for TbMnO$_3$ is found to be in good agreement with the measured data [8] with sharp peaks at 27 K and 42 K observed by Kimura et al; these anomalies are associated with the incommensurate-commensurate lock-in transition at $T_{lock}$ = 27 K and the antiferromagnetic order of the Mn$^{3+}$ ions at $T_N$ = 42 K, respectively.

An inspection of the Fig. 3(b) shows that the computed specific heat variation for DyMnO$_3$ is close to the experimental data [8]. It is also seen from this figure that the experimental results exhibit double peaks at 20 and 40 K, which are due to the occurrence of the weak ferromagnetic (FM) component and antiferromagnetic (AF) ordering. Further, Fig. 3 (b) reveals that our results vary systematically with temperature from 100 to 1000 K and found to follow the similar trend as is exhibited by the experimental data [8] from 2 to 100 K. For clarity, this feature is also depicted in the inset. The experimental curve for HoMnO$_3$, shown in Fig. 3 (c), shows that there are magnetic transitions below 60 K [10]. Since our computations take account of only the phonon contribution to the specific heat, therefore, these anomalies, arising due to the magnetic interactions, are not revealed from our calculated specific heat results in all the three manganites.

### 3.6 DEBYE TEMPERATURES

The calculated values of Debye temperatures ($\theta_D$) 574 K (TbMnO$_3$), 574 K (DyMnO$_3$) and 577 K (HoMnO$_3$) at 300 K are in fair agreement with the corresponding experimental values (300–600 K) [16] often found in perovskite manganites. These values of $\theta_D$ have been plotted against the temperature in Fig. 4 for $R$MnO$_3$ (R=Tb, Dy, Ho) in the range of temperature up to 1000 K.

### 3.7 PHONON DISPERSION CURVES

The phonon dispersion curves (PDCs) for $R$MnO$_3$ (R=Tb, Dy, Ho) along the three high symmetry directions Σ, Δ and Λ in the Brillouin zone have been computed using the present shell model and DISPR program [24]. The computed PDCs are depicted in Figs. 5-7. The modes belonging to the different representations are plotted separately in the same figure. With 20 atoms in the unit cell, the phonon dispersion curves have 60 phonon branches along each symmetry direction. At the zone boundary points X, Y and Z, the branches are degenerate in pairs due to the high symmetry at these points. It is interesting to note that the PDCs along Σ$_2$, Δ$_2$ and Λ$_2$ directions from Γ to X, Y and Z points have exhibited some imaginary frequency values in all three TbMnO$_3$. DyMnO$_3$ and HoMnO$_3$ manganites as shown in Figs. 5-7. These features of imaginary phonon modes indicate the possibility of occurrence of phase transformation in these manganites. From the phonon dispersion curves, it is noticed that for o-RMnO$_3$, one A$_u$ mode at the Γ point is not a stable mode.

## 3.8 GIBBS FREE ENERGY AND PHASE COEXISTENCE

The dynamic instabilities seen in the orthorhombic phase and the known existence of the hexagonal phase for $R$MnO$_3$ (R=Tb, Dy, Ho) suggests that the criteria of polymorph formation should be of an energetic nature correlating with a small energy difference. Calculation of the Gibbs free energies in these two phases for the different compounds was carried out using the limited data available on the structure of the hexagonal phases [12-15], and by transferring the shell model parameters from the orthorhombic phase. The Gibbs free energy for the two phases is compared in the Figs. 8 and 9, and the energy difference is seen to be less than 2 kJ mol$^{-1}$. This is a clear indication leading to the coexistence of orthorhombic and hexagonal phases in RMnO$_3$ (R=Tb, Dy, Ho).

## 4. SUMMARY

In summary, the phonon dynamics and thermodynamical properties have been investigated using the shell model to reveal (i) the phonon properties of RMnO$_3$ (R=Tb, Dy, Ho) perovskite orthomanganites and their manifestations in thermodynamic quantities like Debye temperature, specific heat, etc., and (ii) the computation of the elastic constants of structurally complex materials from atomistic approach involving semiempirical interatomic potential. Our calculated crystal structure parameters and Raman frequencies are found to be in good agreement with the available experimental data. The specific heat, lattice dynamical and thermodynamical calculations of structurally complex manganites emphasize the role of theoretical models for the investigation of the panorama of macroscopic physical properties. The results obtained from the present shell model reproduce well the observed data available.

On the basis of an overall description, it may be concluded that the present shell model is successful and represents proper approach to describe the phonon dynamical and related properties of the manganite family of perovskites. This model has revealed the structural, elastic, vibrational, and thermodynamic properties of RMnO$_3$ (R=Tb, Dy, Ho) manganites fairly well implying that the interatomic interactions in these manganites are predominantly ionic. Some of the results on the elastic constants and the phonon-dispersion curves are, probably, reported for the first time and these results are although only of academic interest at present but they will serve as a guide to the experimental workers in future. Also, the studies about optical phonons may greatly help in further research of the doping and temperature dependence of the phonon modes in R$_{1-x}$A$_x$MnO$_3$ (R = Tb, Dy, Ho; A= Ca, Sr, Ba) and of their interplay with the colossal magnetoresistance (CMR) effect. The present results will be very useful in the planning, execution and analysis of various experiments by investigators.

**ACKNOWLEDGEMENTS**

Renu Choithrani is grateful to the Council of Scientific and Industrial Research (CSIR) and to the University Grants Commission (UGC), Government of India, New Delhi for the award of Senior Research Fellowship (SRF) and for providing the financial support.

**CAPTION OF FIGURES**

| | |
|---|---|
| FIG. 1 | The calculated long-wavelength (infrared active (LO—squares; TO—circles) and silent (upright triangles)) phonon frequencies and mode assignments for orthorhombic manganites $RMnO_3$ (R=Tb, Dy, Ho) (1 meV=8.059 cm$^{-1}$=0.242 THz) |
| FIG. 2 | The calculated partial (for the various atoms) and total phonon density of states for $RMnO_3$ (R=Tb, Dy, Ho). All the phonon spectra have been normalized to unity (1meV= 8.059cm$^{-1}$ = 0.242THz). |
| FIG. 3 | The specific heat of $RMnO_3$ (R=Tb, Dy or Ho) as a function of temperature, where solid line (─) and the squares (□) are the present model calculation and the experimental [8, 10] values, respectively. |
| FIG. 4 | The variation of Debye temperature of $RMnO_3$ (R=Tb, Dy or Ho) with temperature as obtained from shell model calculations. |
| FIG.5 | The computed phonon-dispersion curves for $TbMnO_3$ along the high symmetry $\sum$, $\Delta$ and $\Lambda$ directions. Group-theoretical representations are indicated on top of the figure. $\Gamma$ is the Brillouin zone center and X, Y and Z are the zone boundary points (1 meV = 8.059 cm−1 = 0.242 THz). |
| FIG.6 | The computed phonon-dispersion curves for $DyMnO_3$ along the high symmetry $\sum$, $\Delta$ and $\Lambda$ directions. Group-theoretical representations are indicated on top of the figure. $\Gamma$ is the Brillouin zone center and X, Y and Z are the zone boundary points (1 meV = 8.059 cm$^{-1}$ = 0.242 THz). |
| FIG. 7 | The computed phonon-dispersion curves for $HoMnO_3$ along the high symmetry $\sum$, $\Delta$ and $\Lambda$ directions. Group-theoretical representations are indicated on top of the figure. $\Gamma$ is the Brillouin zone center and X, Y and Z are the zone boundary points (1 meV = 8.059 cm$^{-1}$ = 0.242 THz). |
| FIG. 8 | Calculated Gibbs free energy for the orthorhombic and hexagonal phases of $RMnO_3$ (R=Tb,Dy,Ho), as a function of temperature. |
| FIG. 9 | Calculated Gibbs free energy for the orthorhombic and hexagonal phases of $RMnO_3$ (R=Tb,Dy,Ho) as a function of pressure. The solid (orthorhombic phase) and dashed (hexagonal phase) lines are just guides to the eye. |

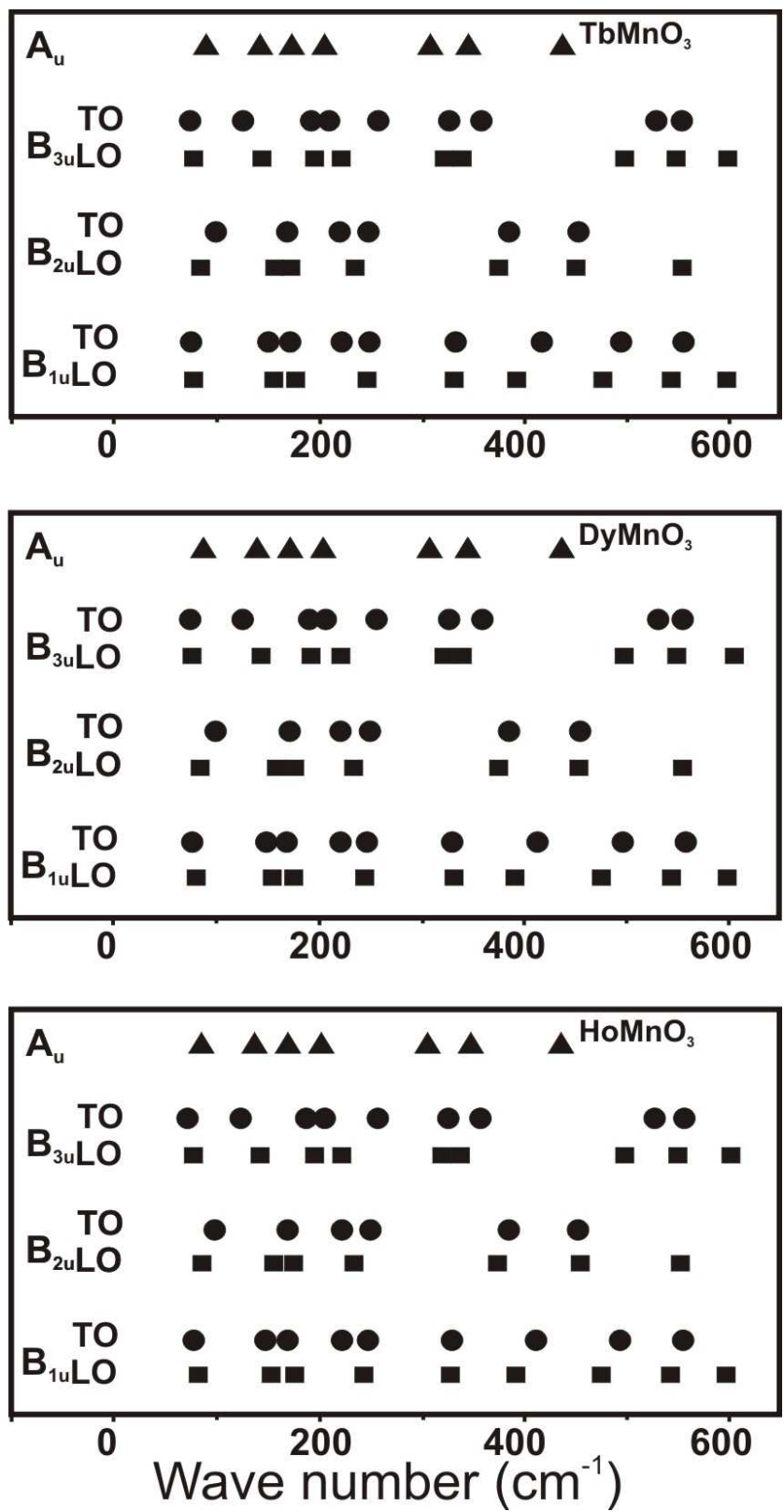

**Fig. 1**

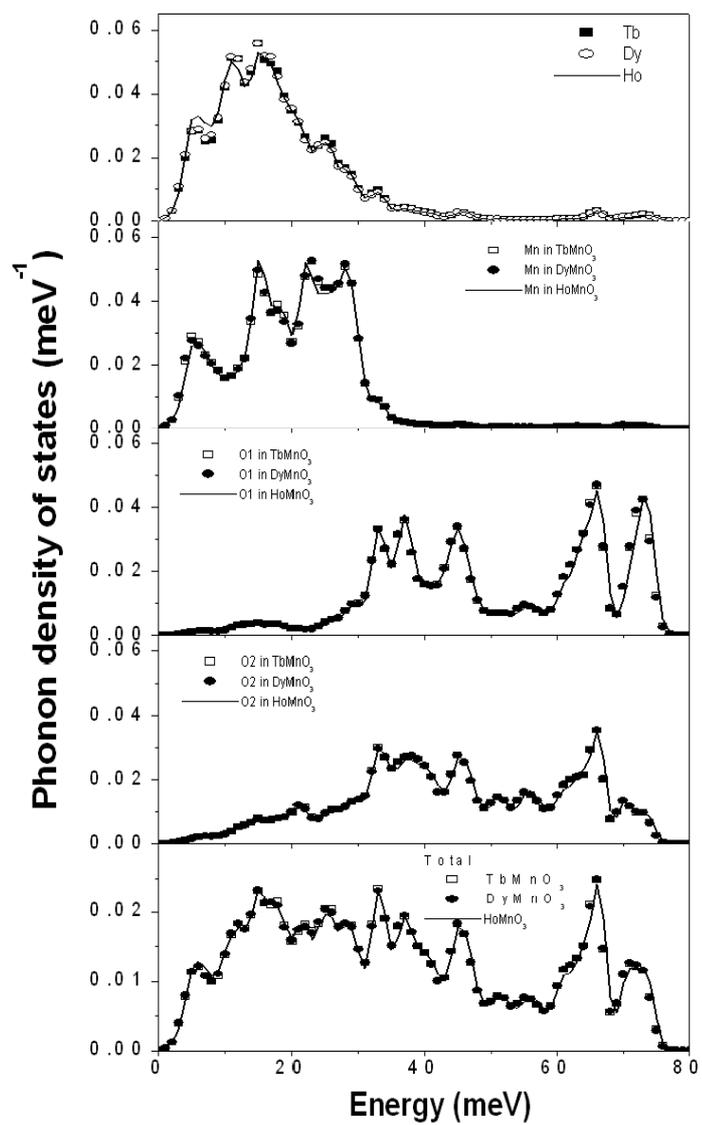

**Fig. 2**

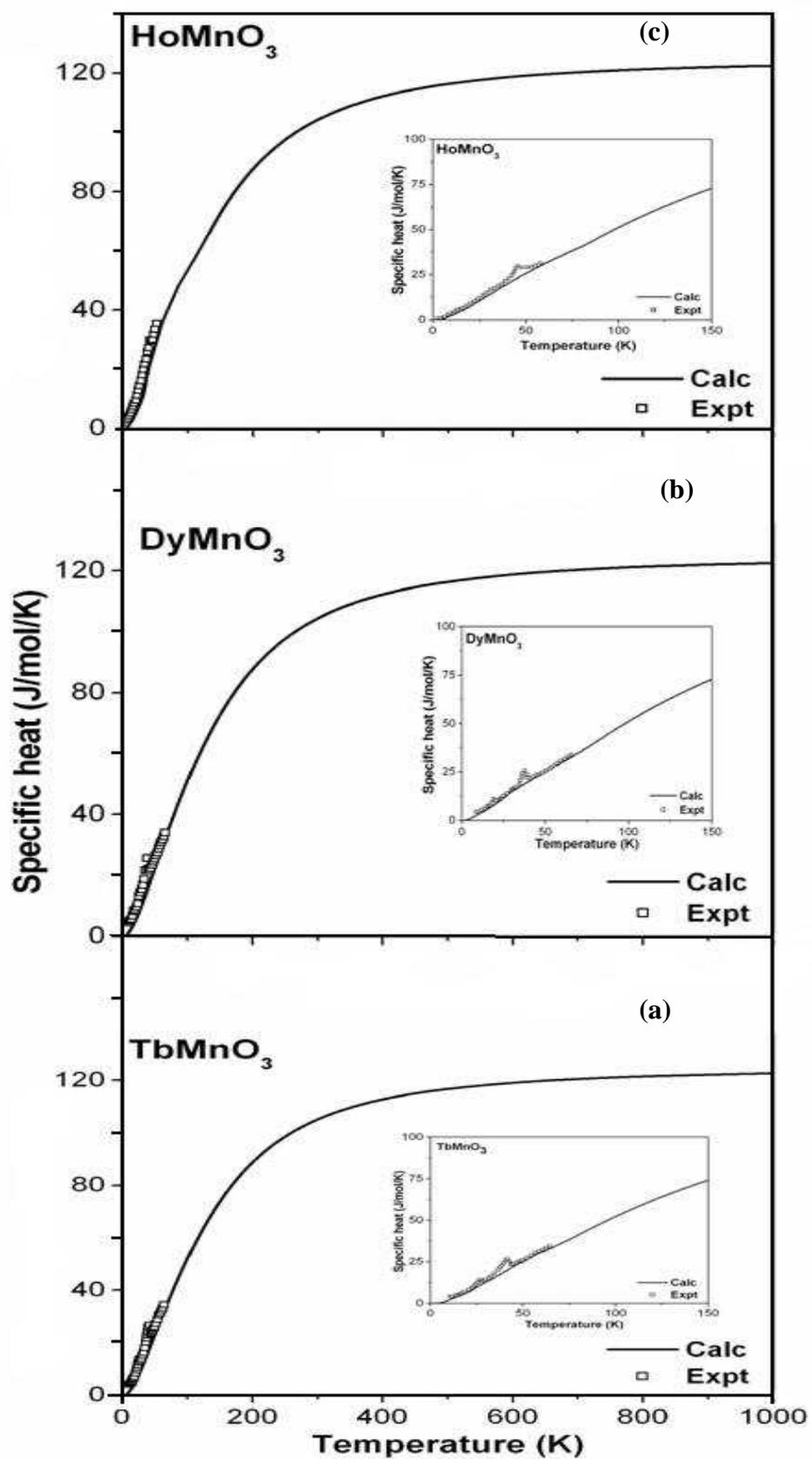

**Fig. 3**

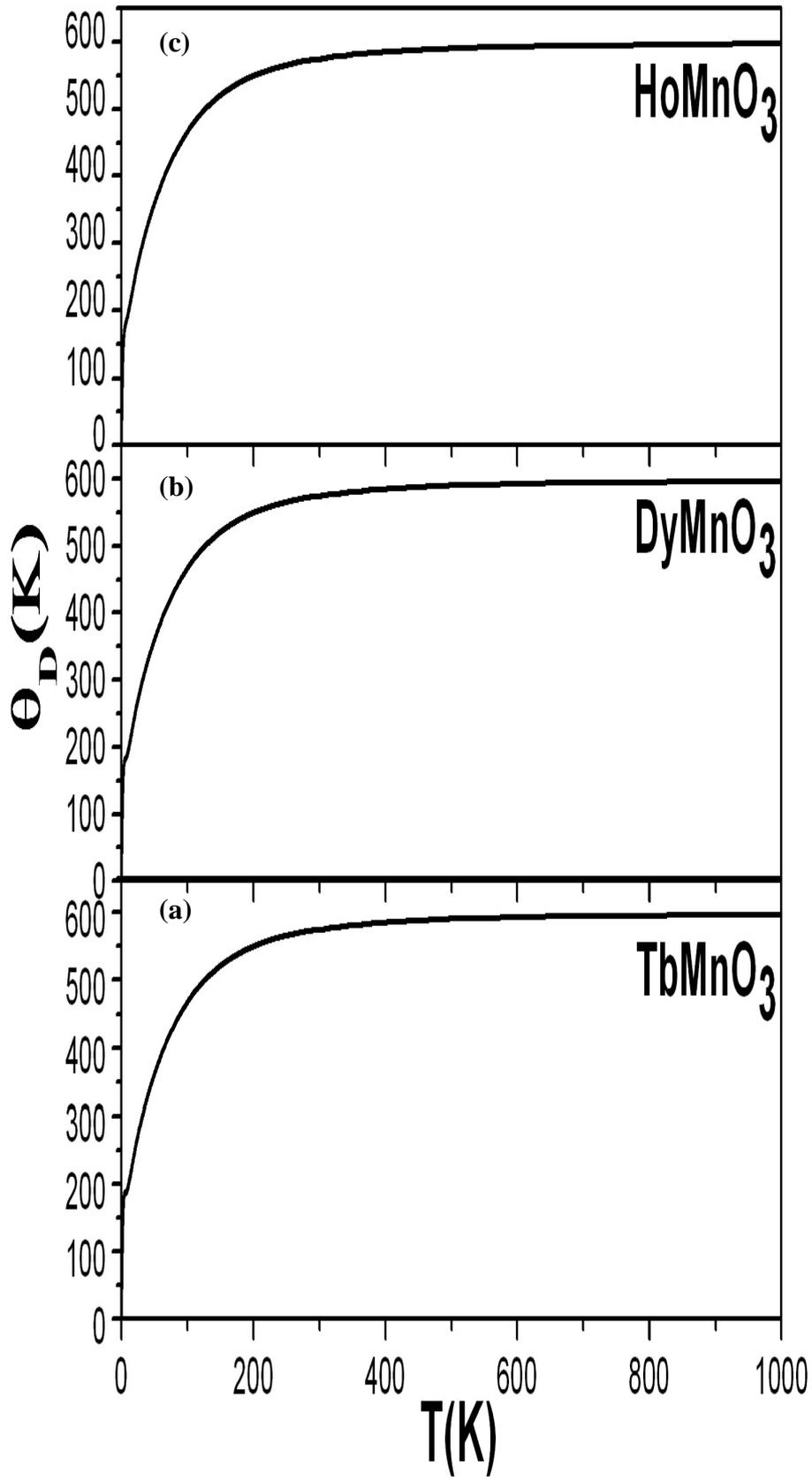

Fig. 4

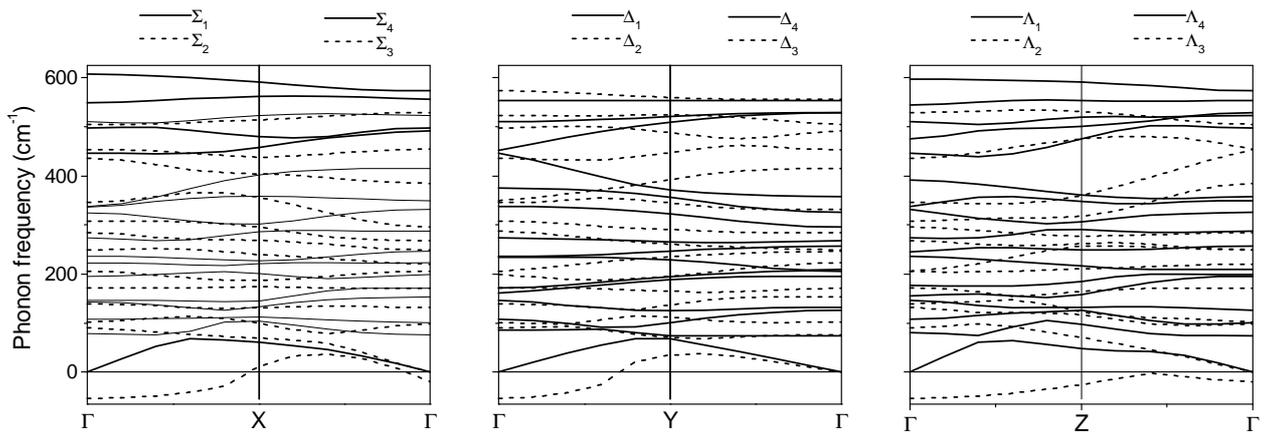

**Fig. 5 PDR of TbMnO₃**

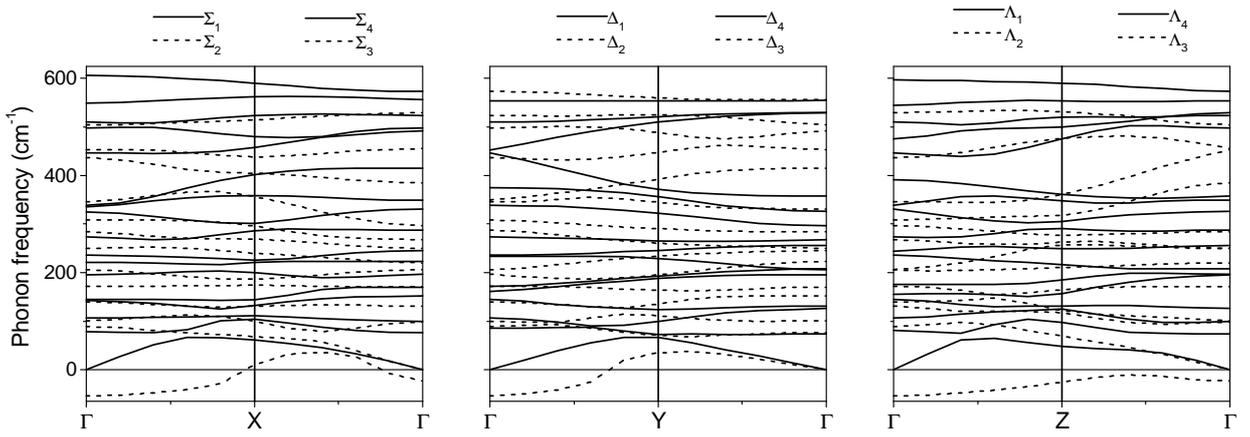

**Fig. 6. PDR of DyMnO₃**

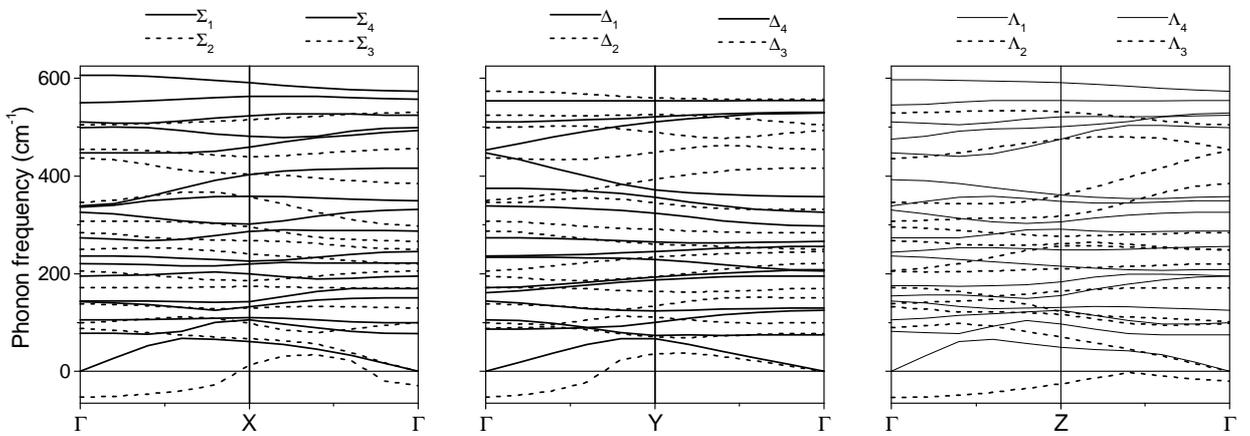

**Fig. 7. PDR of HoMnO₃**

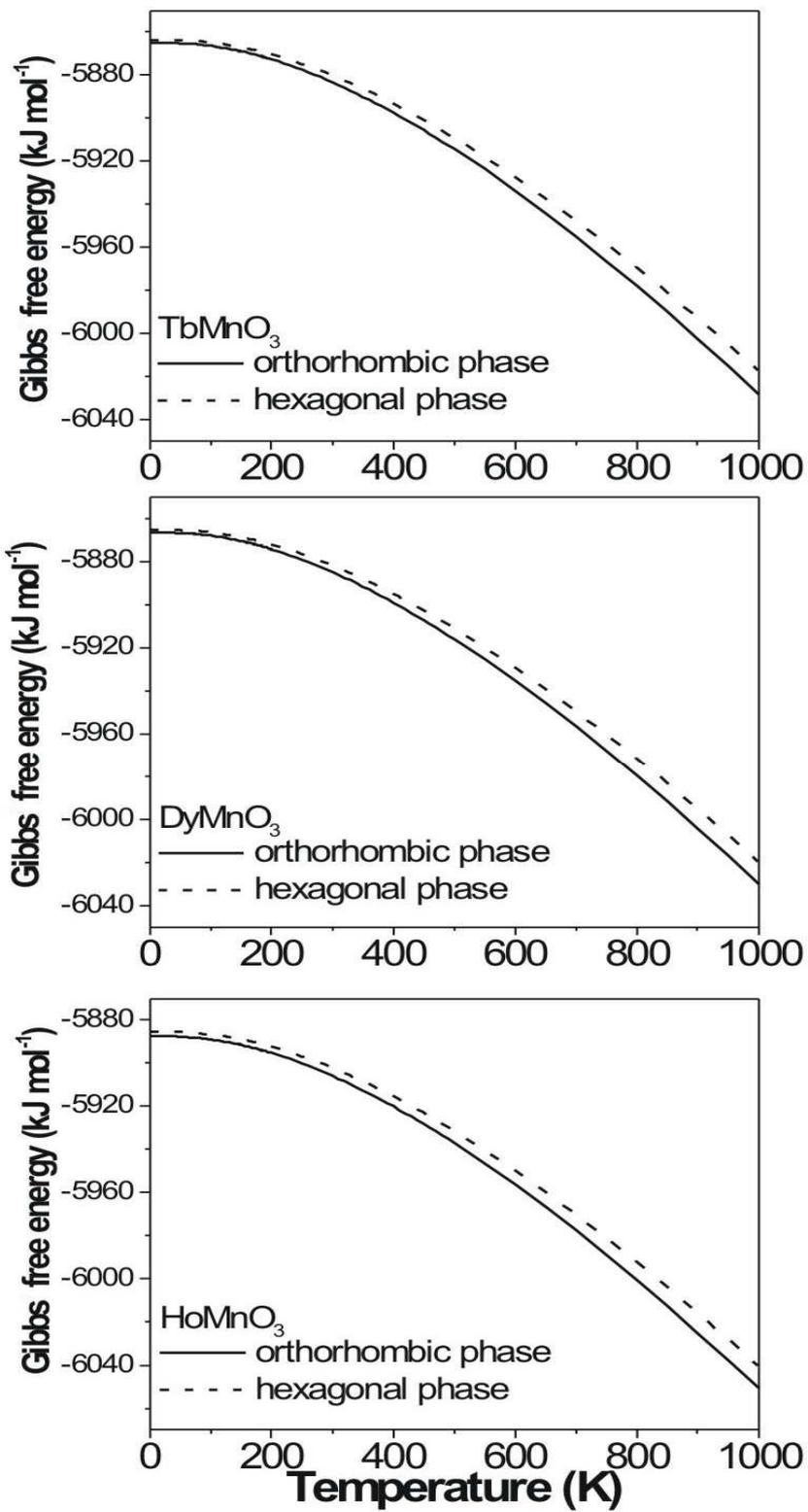

**Fig. 8.**

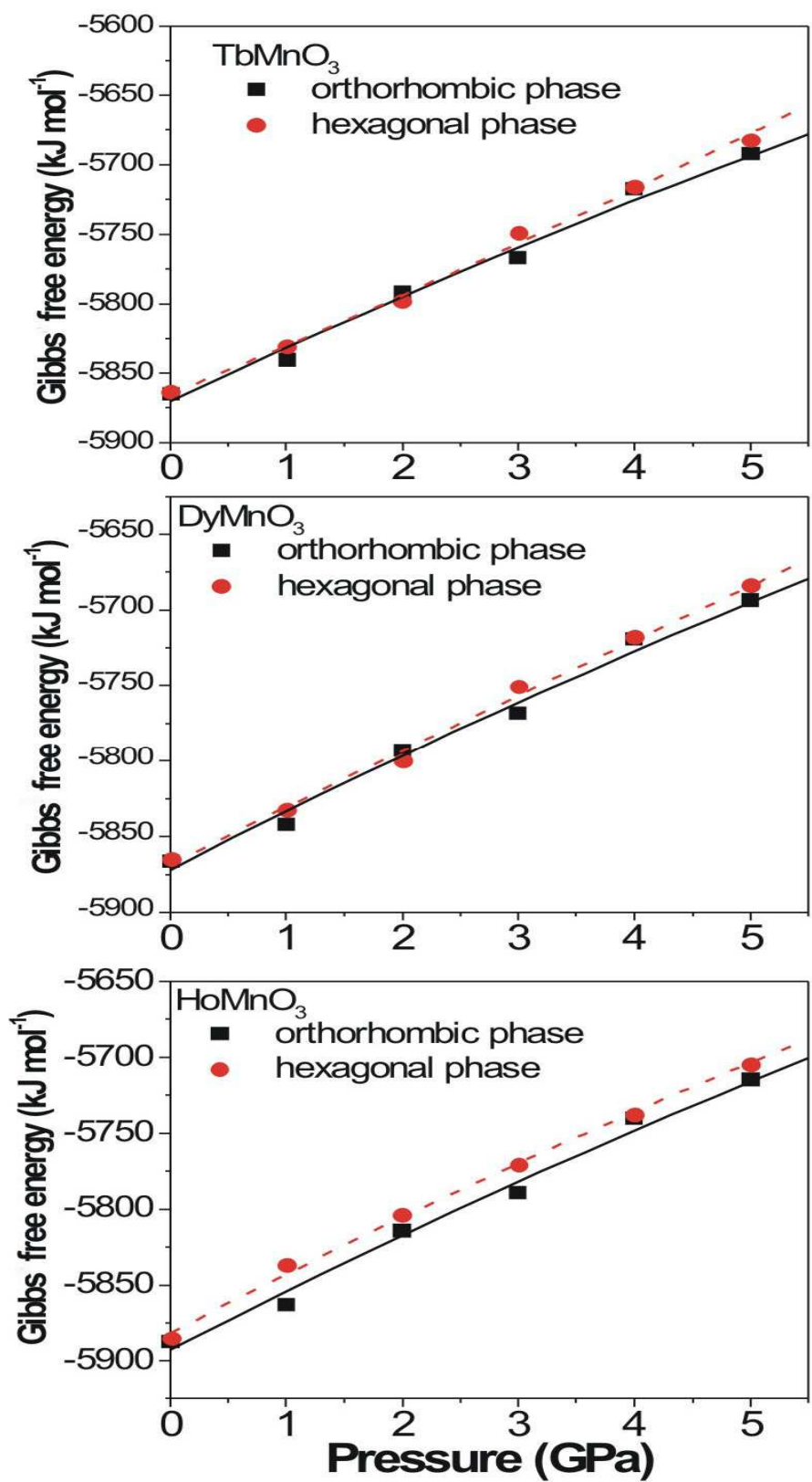

**Fig. 9**